\documentclass[]{IEEEtran}

\usepackage{hyperref}    %
\usepackage{url}      %
\usepackage{booktabs}    %
\usepackage{amsfonts}    %
\usepackage{nicefrac}    %
\usepackage{microtype}   %
\usepackage{graphicx}
\usepackage{subcaption}
\usepackage{soul}
\usepackage{amsmath}
\usepackage{amsthm} %
\usepackage{amssymb} %
\usepackage{palatino} %
\usepackage{algorithm}
\usepackage[noend]{algpseudocode}
\usepackage{multirow}
\usepackage{tablefootnote}
\usepackage{threeparttable}

\usepackage{color}

\usepackage{footnote}
\makesavenoteenv{tabular}
\makesavenoteenv{table}

\newcommand{\furl}[1]{\footnote{\url{https://#1}}}

\theoremstyle{definition}

\begin{document}

\title{Suicidal Ideation Detection: A Review of Machine Learning Methods and Applications}

\author{Shaoxiong Ji,
	Shirui Pan,~\IEEEmembership{Member,~IEEE,}
    Xue Li,\\
    Erik Cambria,~\IEEEmembership{Senior Member,~IEEE,}
    Guodong Long,
    and~Zi Huang%
\IEEEcompsocitemizethanks{\IEEEcompsocthanksitem S. Ji is with Aalto University, Finland and The University of Queensland, Australia. E-mail:~shaoxiong.ji@aalto.fi
\IEEEcompsocthanksitem X. Li, and Z. Huang are with The University of Queensland, Australia.
E-mail:~\{xueli;~huang\}@itee.uq.edu.au
\IEEEcompsocthanksitem S. Pan is with Monash University, Australia. E-mail:~shirui.pan@monash.edu
\IEEEcompsocthanksitem E. Cambria is with Nanyang Technological University, Singapore. E-mail:~cambria@ntu.edu.sg
\IEEEcompsocthanksitem G. Long is with University of Technology Sydney, Australia. E-mail:~guodong.long@uts.edu.au}
}

\IEEEtitleabstractindextext{%
\begin{abstract}
Suicide is a critical issue in modern society. Early detection and prevention of suicide attempts should be addressed to save people's life. 
Current suicidal ideation detection methods include clinical methods based on the interaction between social workers or experts and the targeted individuals and machine learning techniques with feature engineering or deep learning for automatic detection based on online social contents. 
This paper is the first survey that comprehensively introduces and discusses the methods from these categories.
Domain-specific applications of suicidal ideation detection are reviewed according to their data sources, i.e., questionnaires, electronic health records, suicide notes, and online user content. Several specific tasks and datasets are introduced and summarized to facilitate further research. 
Finally, we summarize the limitations of current work and provide an outlook of further research directions.
\end{abstract}

}

\maketitle

\IEEEdisplaynontitleabstractindextext

\IEEEpeerreviewmaketitle

\section{Introduction}
\label{sec::introduction}

\IEEEPARstart{M}{ental} health issues, such as anxiety and depression, are becoming increasingly concerned in modern society, as they turn out to be especially severe in developed countries and emerging markets. Severe mental disorders without effective treatment can turn to suicidal ideation or even suicide attempts.
Some online posts contain much negative information and generate problematic phenomena such as cyberstalking and cyberbullying. Consequences can be severe and risky since such lousy information is often engaged in some form of social cruelty, leading to rumors or even mental damage. Research shows that there is a link between cyberbullying and suicide~\cite{hinduja2010bullying}. Victims overexposed to too many negative messages or events may become depressed and desperate; even worse, some may commit suicide. 

The reasons that people commit suicide are complicated. People with depression are highly likely to commit suicide, but many without depression can also have suicidal thoughts~\cite{joo2016death}. 
According to the American Foundation for Suicide Prevention (AFSP), suicide factors fall under three categories: health factors, environmental factors, and historical factors~\cite{vioules2018detection}. Ferrari et al.~\cite{ferrari2014burden} found that mental health issues and substance use disorders are attributed to the factors of suicide. 
O'Connor and Nock~\cite{o2014psychology} conducted a thorough review of the psychology of suicide and summarized psychological risks as personality and individual differences, cognitive factors, social factors, and negative life events.

Suicidal Ideation Detection (SID) determines whether the person has suicidal ideation or thoughts by given tabular data of a person or textual content written by a person.
Due to the advances in social media and online anonymity, an increasing number of individuals turn to interact with others on the Internet. Online communication channels are becoming a new way for people to express their feelings, suffering, and suicidal tendencies. Hence, online channels have naturally started to act as a surveillance tool for suicidal ideation, and mining social content can improve suicide prevention~\cite{lopez2019mining}. Strange social phenomena are emerging, e.g., online communities reaching an agreement on self-mutilation and copycat suicide. For example, a social network phenomenon called the ``Blue Whale Game''\furl{thesun.co.uk/news/worldnews/3003805} in 2016 uses many tasks (such as self-harming) and leads game members to commit suicide in the end. Suicide is a critical social issue and takes thousands of lives every year. Thus, it is necessary to detect suicidality and to prevent suicide before victims end their life. Early detection and treatment are regarded as the most effective ways to prevent potential suicide attempts. 

Potential victims with suicidal ideation may express their thoughts of committing suicide in fleeting thoughts, suicide plans, and role-playing. 
Suicidal ideation detection is to find out these risks of intentions or behaviors before tragedy strikes.
A meta-analysis conducted by McHugh et al.~\cite{mchugh2019association} shown statistical limitations of ideation as a screening tool, but also pointed out that people's expression of suicidal ideation represents their psychological distress. 
Effective detection of early signals of suicidal ideation can identify people with suicidal thoughts and open a communication portal to let social workers mitigate their mental issues. 
The reasons for suicide are complicated and attributed to a complex interaction of many factors~\cite{o2014psychology, kassen2019behavioral}. 
To detect suicidal ideation, many researchers conducted psychological and clinical studies~\cite{venek2017adolescent} and classified responses of questionnaires~\cite{delgado2011improving}. Based on their social media data, artificial intelligence (AI) and machine learning techniques can predict people's likelihood of suicide~\cite{liu2019socinf}, which can better understand people's intentions and pave the way for early intervention. Detection on social content focuses on feature engineering~\cite{odea2015detecting, shing2018expert}, sentiment analysis~\cite{ren2016examining, yue2018survey}, and deep learning~\cite{benton2017multi, ji2018supervised, ji2019detecting}. 
Those methods generally require heuristics to select features or design artificial neural network architectures for learning rich representation. The research trend focuses on selecting more useful features from people's health records and developing neural architectures to understand the language with suicidal ideation better.

Mobile technologies have been studied and applied to suicide prevention, for example, the mobile suicide intervention application iBobbly~\cite{tighe2017ibobbly} developed by the Black Dog Institute\furl{blackdoginstitute.org.au/research/digital-dog/programs/ibobbly-app}. Many other suicide prevention tools integrated with social networking services have also been developed, including Samaritans Radar\furl{samaritans.org/about-samaritans/research-policy/internet-suicide/samaritans-radar} and Woebot\furl{woebot.io}. The former was a Twitter plugin that was later discontinued because of privacy issues. For monitoring alarming posts. The latter is a Facebook chatbot based on cognitive behavioral therapy and natural language processing (NLP) techniques for relieving people's depression and anxiety. 

Applying cutting-edge AI technologies for suicidal ideation detection inevitably comes with privacy issues~\cite{de2018ethics} and ethical concerns~\cite{mckernan2018protecting}. 
Linthicum et al.~\cite{linthicum2019machine} put forward three ethical issues, including the influence of bias on machine learning algorithms, the prediction on time of suicide act, and ethical and legal questions raised by false positive and false negative prediction. 
It is not easy to answer ethical questions for AI as these require algorithms to reach a balance between competing values, issues, and interests~\cite{de2018ethics}.

\begin{figure*}[!ht]
\begin{center}
\includegraphics[width=0.85\textwidth]{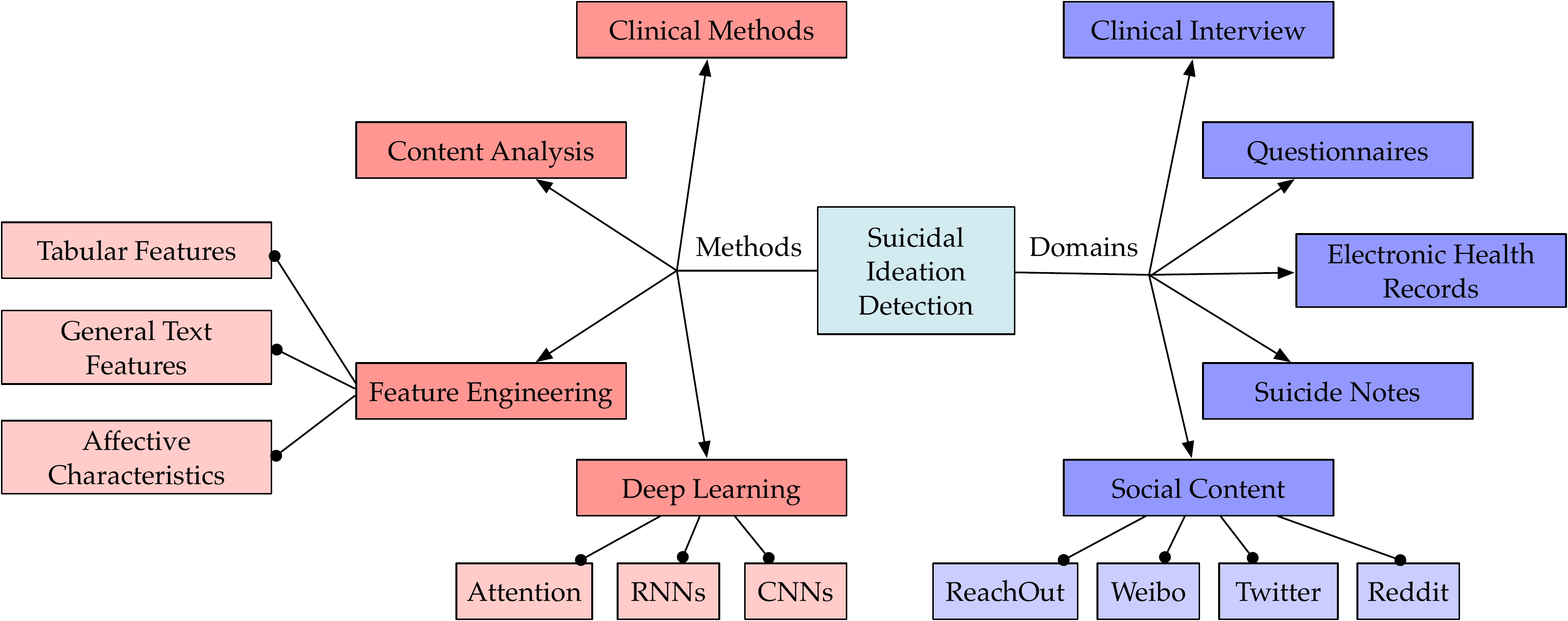}
\caption{The categorization of suicide ideation detection: methods and domains. The left part represents method categorization, while the right part shows the categories of domains. The arrow and solid point indicate subcategories.}
\label{fig:categorization}
\end{center}
\end{figure*}

AI has been applied to solve many challenging social problems. Detection of suicidal ideation with AI techniques is one of the potential applications for social good and should be addressed to improve people's wellbeing meaningfully. 
The research problems include feature selection on tabular and text data and representation learning on natural language. Many AI-based methods have been applied to classify suicide risks. However, there remain some challenges. There is a limited number of benchmarks for training and evaluating suicidal ideation detection. AI-powered models sometimes learn statistical clues, but fail to understand people's intention. Moreover, many neural models are lack of interpretability.
This survey reviews suicidal ideation detection methods from the perspective of AI and machine learning and specific domain applications with social impact. The categorization from these two perspectives is shown in Fig.~\ref{fig:categorization}. 
This paper provides a comprehensive review of the increasingly important field of suicidal ideation detection with machine learning methods. It proposes a summary of current research progress and an outlook of future work.
The contributions of our survey are summarized as follows.
\begin{itemize}
\item To the best of our knowledge, this is the first survey that conducts a comprehensive review of suicidal ideation detection, its methods, and its applications from a machine learning perspective.
\item We introduce and discuss the classical content analysis and modern machine learning techniques, plus their application to questionnaires, EHR data, suicide notes, and online social content.
\item We enumerate existing and less explored tasks and discuss their limitations. We also summarize existing datasets and provide an outlook of future research directions in this field. 
\end{itemize}

The remainder of the paper is organized as follows: methods and applications are introduced and summarized in Section~\ref{sec:methods} and Section~\ref{sec:application}, respectively; Section~\ref{sec:tasks} enumerates specific tasks and some datasets; finally, we have a discussion and propose some future directions in Section~\ref{sec:discussion}.

\section{Methods~and~Categorization}
\label{sec:methods}

Suicide detection has drawn the attention of many researchers due to an increasing suicide rate in recent years and has been studied extensively from many perspectives. 
The research techniques used to examine suicide also span many fields and methods, for example, clinical methods with patient-clinic interaction~\cite{venek2017adolescent} and automatic detection from user-generated content (mainly text)~\cite{odea2015detecting, ji2018supervised}. Machine learning techniques are widely applied for automatic detection.

Traditional suicide detection relies on clinical methods, including self-reports and face-to-face interviews.
Venek et al.~\cite{venek2017adolescent} designed a five-item ubiquitous questionnaire for the assessment of suicidal risks and applied a hierarchical classifier on the patients' response to determine their suicidal intentions. Through face-to-face interaction, verbal and acoustic information can be utilized. Scherer~\cite{scherer2013investigating} investigated the prosodic speech characteristics and voice quality in a dyadic interview to identify suicidal and non-suicidal juveniles. Other clinical methods examine resting state heart rate from converted sensing signals~\cite{sikander2016predicting}, classify functional magnetic resonance imaging-based neural representations of death- and life-related words~\cite{just2017machine}, and event-related instigators converted from EEG signals~\cite{jiang2015erp}.
Another aspect of clinical treatment is the understanding of the psychology behind suicidal behavior~\cite{o2014psychology}, which, however, relies heavily on the clinician's knowledge and face-to-face interaction. Suicide risk assessment scales with clinical interview can reveal informative cues for predicting suicide~\cite{lotito2015review}. 
Tan et al.~\cite{tan2017designing} conducted an interview and survey study in Weibo, a Twitter-like service in China, to explore the engagement of suicide attempters with intervention by direct messages. 

\subsection{Content~Analysis}
\label{sec:content-analysis}
Users' post on social websites reveals rich information and their language preferences. Through exploratory data analysis on the user-generated content can have an insight into language usage and linguistic clues of suicide attempters. The detailed analysis includes lexicon-based filtering, statistical linguistic features, and topic modeling within suicide-related posts.

Suicide-related keyword dictionary and lexicon are manually built to enable keyword filtering~\cite{huang2007hunting, varathan2014suicide} and phrases filtering~\cite{jashinsky2014tracking}. Suicide-related keywords and phrases include ``kill'', ``suicide'', ``feel alone", ``depressed'', and ``cutting myself''. 
Vioul\`{e}s et al.~\cite{vioules2018detection} built a point-wise mutual information symptom lexicon using an annotated Twitter dataset. Gunn and Lester~\cite{gunn2015twitter} analyzed posts from Twitter in the 24 hours before the death of a suicide attempter. Coppersmith et al.~\cite{coppersmith2015quantifying} analyzed the language usage of data from the same platform.
Suicidal thoughts may involve strong negative feelings, anxiety, and hopelessness, or other social factors like family and friends. Ji et al.~\cite{ji2018supervised} performed word cloud visualization and topics modeling over suicide-related content and found that suicide-related discussion covers personal and social issues. Colombo et al.~\cite{colombo2016analysing} analyzed the graphical characteristics of connectivity and communication in the Twitter social network. Coppersmith et al.~\cite{coppersmith2016exploratory} provided an exploratory analysis of language patterns and emotions on Twitter. 
Other methods and techniques include Google Trends analysis for suicide risk monitoring~\cite{solano2016google}, the reply bias assessment through linguistic clues~\cite{huang2016online}, human-machine hybrid method for analysis of the language effect of social support on suicidal ideation risk~\cite{de2017language}, social media content detection, and speech patterns analysis~\cite{larsen2015use}.

\subsection{Feature~Engineering}
\label{sec:feature-engineering}
The goal of text-based suicide classification is to determine whether candidates, through their posts, have suicidal ideations. 
Machine learning methods and NLP have also been applied in this field. 

\subsubsection{Tabular~Features}
Tabular data for suicidal ideation detection consist of questionnaire responses and structured statistical information extracted from websites. Such structured data can be directly used as features for classification or regression. 
Masuda et al.~\cite{masuda2013suicide} applied logistic regression to classify suicide and control groups based on users' characteristics and social behavior variables. The authors found variables such as community number, local clustering coefficient, and homophily have a more substantial influence on suicidal ideation in an SNS of Japan. 
Chattopadhyay~\cite{chattopadhyay2007study} applied Pierce Suicidal Intent Scale (PSIS) to assess suicide factors and conducted regression analysis. 
Questionnaires act as a good source of tabular features. Delgado-Gomez et al.~\cite{delgado2012suicide} used the international personality disorder examination screening questionnaire and the Holmes-Rahe social readjustment rating scale.
Chattopadhyay~\cite{chattopadhyay2012mathematical} proposed to apply a multilayer feed-forward neural network, as shown in Fig.~\ref{fig:method-mlp} to classify suicidal intention indicators according to Beck's suicide intent scale. 

\subsubsection{General~Text~Features}
\begin{figure}[htbp]
\begin{center}
\begin{subfigure}[]{0.4\textwidth}
	\includegraphics[height=3cm]{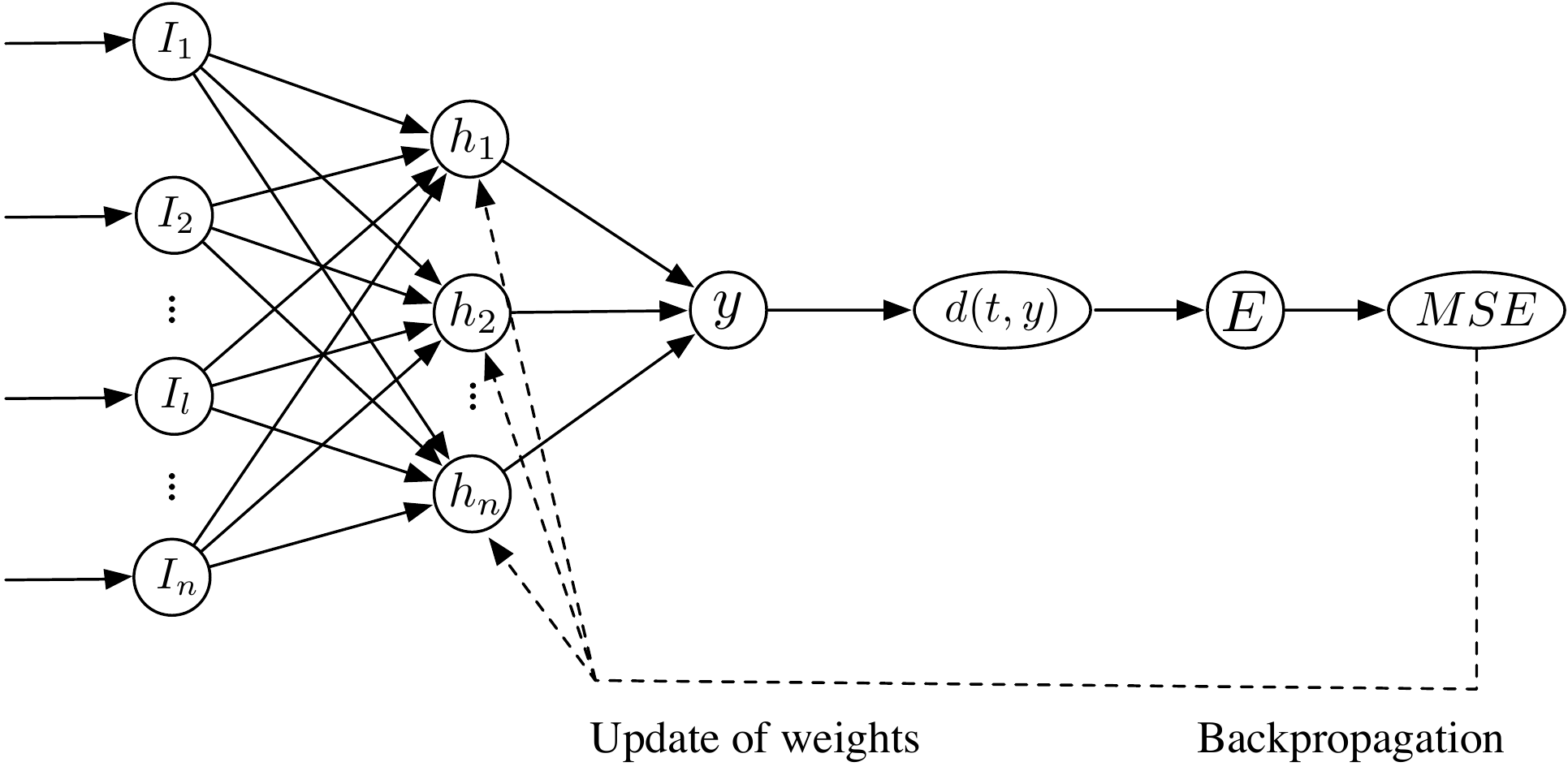}
	\subcaption{Neural network with feature engineering}
	\label{fig:method-mlp}
\end{subfigure}
\qquad
\begin{subfigure}[]{0.5\textwidth}
	\includegraphics[height=3cm]{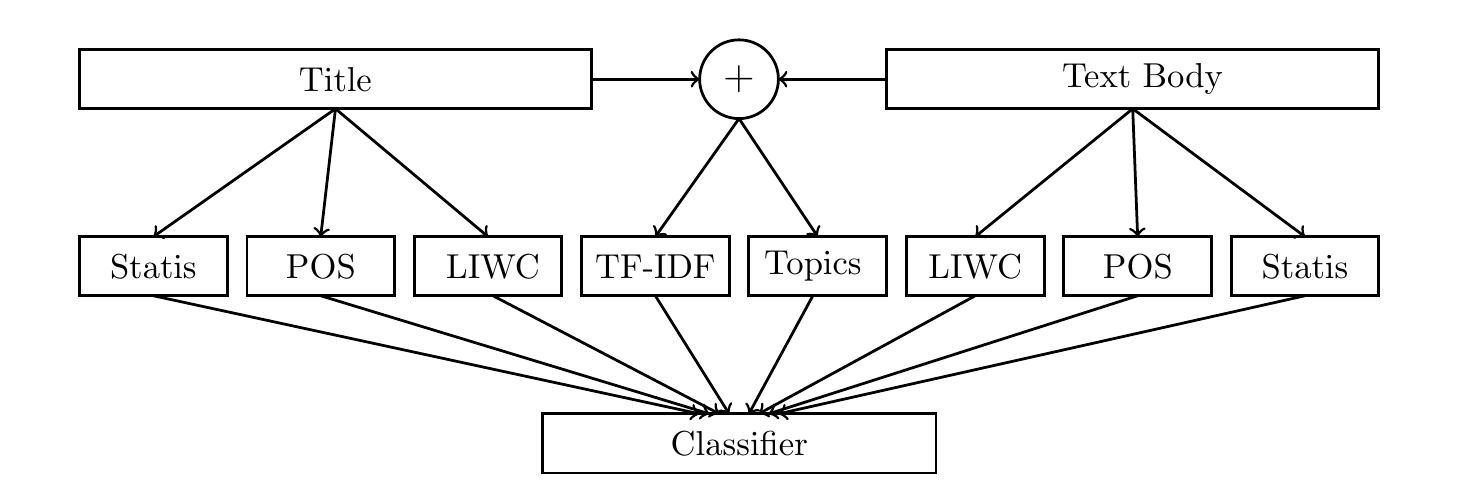}
	\subcaption{Classifier with feature engineering}
	\label{fig:method-ji-2018}
\end{subfigure}
\caption{Illustrations of methods with feature engineering}
\label{fig:method-features}
\end{center}
\end{figure}

Another direction of feature engineering is to extract features from unstructured text. 
The main features consist of N-gram features, knowledge-based features, syntactic features, context features, and class-specific features~\cite{wang2012discovering}. 
Abboute et al.~\cite{abboute2014mining} built a set of keywords for vocabulary feature extraction within nine suicidal topics.
Okhapkina et al.~\cite{okhapkina2017adaptation} built a dictionary of terms about suicidal content. They introduced term frequency-inverse document frequency (TF-IDF) matrices for messages and a singular value decomposition (SVD) for matrices. 
Mulholland and Quinn~\cite{mulholland2013suicidal} extracted vocabulary and syntactic features to build a classifier to predict the likelihood of a lyricist's suicide. Huang et al.~\cite{huang2014detecting} built a psychological lexicon dictionary by extending HowNet (a commonsense word collection) and used a support vector machines (SVM) to detect cybersuicide in Chinese microblogs. The topic model~\cite{huang2015topic} is incorporated with other machine learning techniques for identifying suicide in Sina Weibo. 
Ji et al.~\cite{ji2018supervised} extract several informative sets of features, including statistical, syntactic, linguistic inquiry and word count (LIWC), word embedding, and topic features, and then put the extracted features into classifiers as shown in Fig.~\ref{fig:method-ji-2018}, where four traditional supervised classifiers are compared. Shing et al.~\cite{shing2018expert} extracted several features as a bag of words (BoWs), empath, readability, syntactic features, topic model posteriors, word embeddings, linguistic inquiry and word count, emotion features and mental disease lexicon. 

Models for suicidal ideation detection with feature engineering include SVM~\cite{wang2012discovering}, artificial neural networks (ANN)~\cite{tai2007artificial} and conditional random field (CRF)~\cite{liakata2012three}. Tai et al.~\cite{tai2007artificial} selected several features, including the history of suicide ideation and self-harm behavior, religious belief, family status, mental disorder history of candidates, and their family. 
Pestian et al.~\cite{pestian2010suicide} compared the performance of different multivariate techniques with features of word counts, POS, concepts, and readability scores. Similarly, Ji et al.~\cite{ji2018supervised} compared four classification methods of logistic regression, random forest, gradient boosting decision tree, and XGBoost. Braithwaite et al.~\cite{braithwaite2016validating} validated machine learning algorithms can effectively identify high suicidal risk. 

\subsubsection{Affective~Characteristics}
Affective characteristics are among the most distinct differences between those who attempt suicide and healthy individuals, which has drawn considerable attention from both computer scientists and mental health researchers. 
To detect the emotions in suicide notes, Liakata et al.~\cite{liakata2012three} used manual emotion categories, including anger, sorrow, hopefulness, happiness/peacefulness, fear, pride, abuse, and forgiveness. Wang et al.~\cite{wang2012discovering} employed combined characteristics of both factual (2 categories) and sentimental aspects (13 categories) to discover fine-grained sentiment analysis. Similarly, Pestian et al.~\cite{pestian2010suicide} identified emotions of abuse, anger, blame, fear, guilt, hopelessness, sorrow, forgiveness, happiness, peacefulness, hopefulness, love, pride, thankfulness, instructions, and information.
Ren et al.~\cite{ren2016examining} proposed a complex emotion topic model and applied it to analyze accumulated emotional traits in suicide blogs and to detect suicidal intentions from a blog stream. Specifically, the authors studied accumulate emotional traits, including emotion accumulation, emotion covariance, and emotion transition among eight basic emotions of joy, love, expectation, surprise, anxiety, sorrow, anger, and hate with a five-level intensity.

\subsection{Deep~Learning}
\label{sec:deep-learning}
\begin{figure}[htbp]
\begin{center}
\begin{subfigure}[]{0.4\textwidth}
	\includegraphics[width=\textwidth]{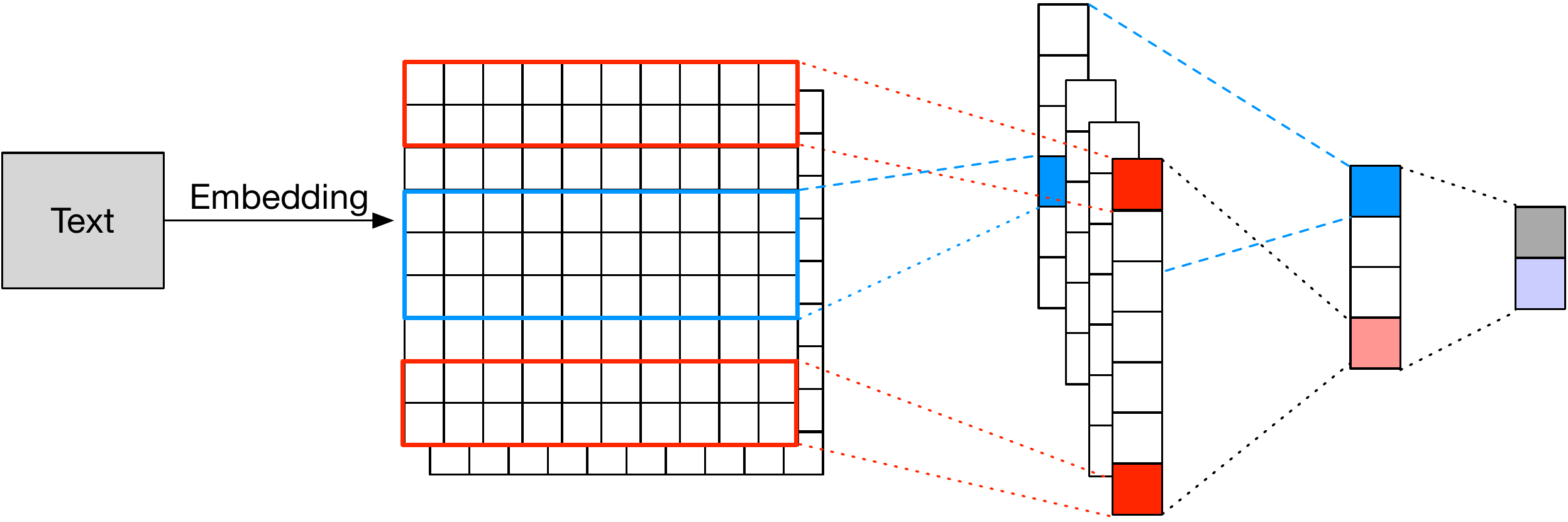}
	\subcaption{CNN}
	\label{fig:CNN}
\end{subfigure}
\qquad
\begin{subfigure}[]{0.4\textwidth}
	\includegraphics[width=\textwidth]{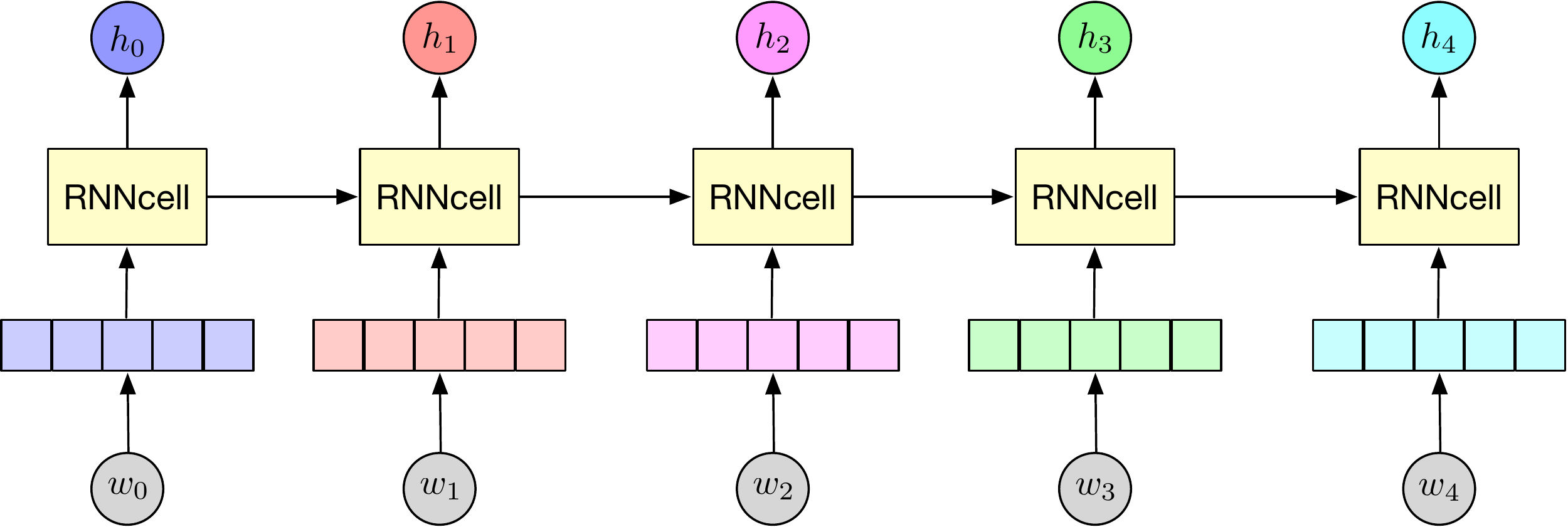}
	\subcaption{RNN}
	\label{fig:RNN}
\end{subfigure}
\qquad
\begin{subfigure}[]{0.4\textwidth}
	\includegraphics[width=\textwidth]{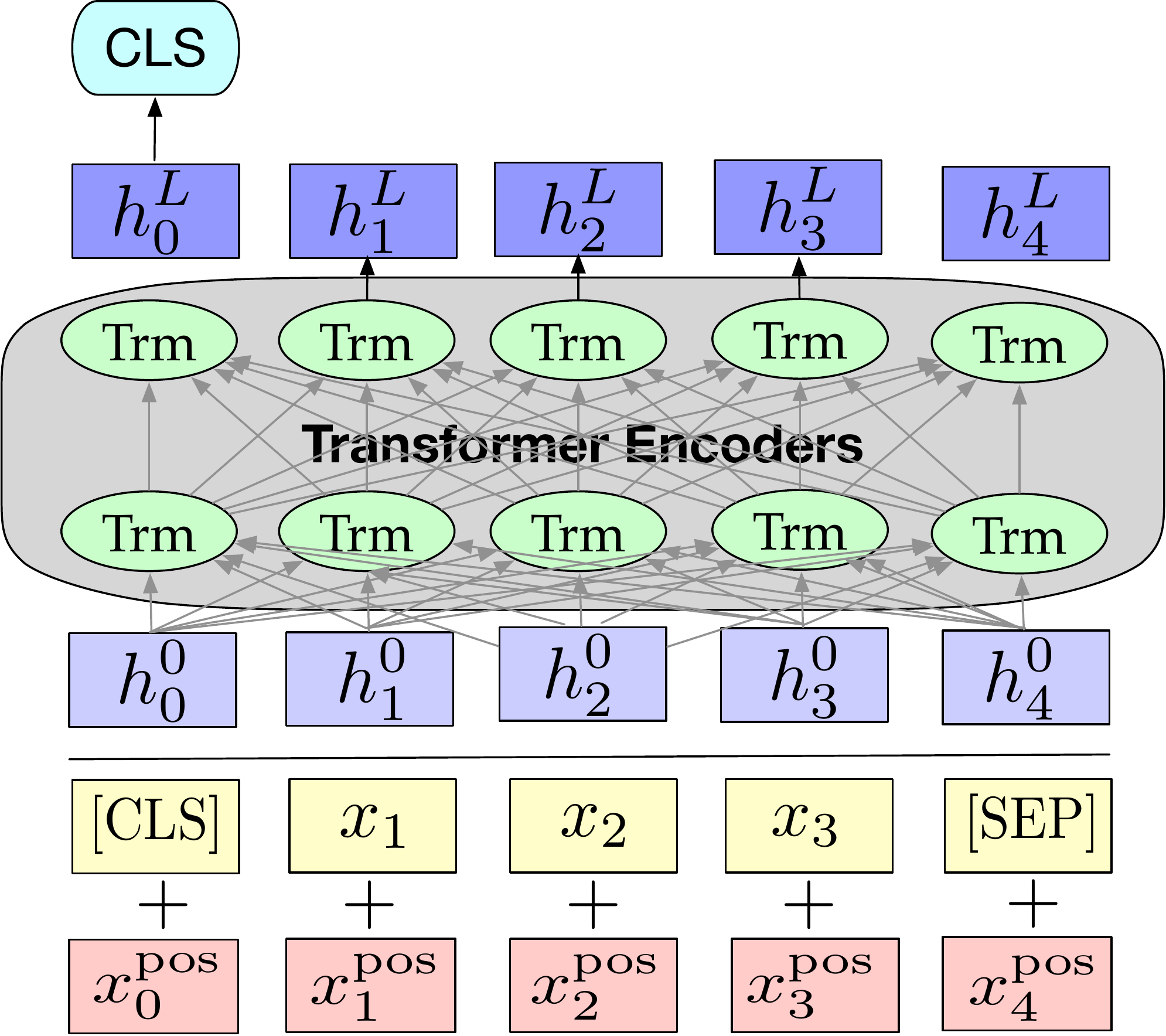}
	\subcaption{BERT}
	\label{fig:BERT}
\end{subfigure}
\caption{Deep neural networks for suicidal ideation detection}
\label{fig:dnn}
\end{center}
\end{figure}

Deep learning has been a great success in many applications, including computer vision, NLP, and medical diagnosis. In the field of suicide research, it is also an important method for automatic suicidal ideation detection and suicide prevention. 
It can effectively learn text features automatically without sophisticated feature engineering techniques. At the same time, some also take extracted features into deep neural networks; for example, Nobles et al.~\cite{nobles2018identification} fed psycholinguistic features and word occurrence into the multilayer perceptron (MLP). Popular deep neural networks (DNNs) include convolutional neural networks (CNNs), recurrent neural networks (RNNs), and bidirectional encoder representations from transformers (BERT), as shown in Fig.~\ref{fig:CNN}, ~\ref{fig:RNN} and~\ref{fig:BERT}. Natural language text is usually embedded into distributed vector space with popular word embedding techniques such as word2vec~\cite{mikolov2013efficient} and GloVe~\cite{pennington2014glove}.
Shing et al.~\cite{shing2018expert} applied user-level CNN with the filter size of 3, 4, and 5 to encode users' posts. 
Long short-term memory (LSTM) network, a popular variant of RNN, is applied to encode textual sequences and then process for classification with fully connected layers~\cite{ji2018supervised}.

Recent methods introduce other advanced learning paradigms to integrate with DNNs for suicidal ideation detection.
Ji et al.~\cite{ji2019knowledge} proposed model aggregation methods for updating neural networks, i.e., CNNs and LSTMs, targeting to detect suicidal ideation in private chatting rooms. However, decentralized training relies on coordinators in chatting rooms to label user posts for supervised training, which can only be applied to minimal scenarios. One possible better way is to use unsupervised or semi-supervised learning methods. Benton et al.~\cite{benton2017multi} predicted suicide attempt and mental health with neural models under the framework of multi-task learning by predicting the gender of users as an auxiliary task. Gaur et al.~\cite{gaur2019knowledge} incorporated external knowledge bases and suicide-related ontology into a text representation and gained an improved performance with a CNN model. Coppersmith et al.~\cite{coppersmith2018natural} developed a deep learning model with GloVe for word embedding, bidirectional LSTM for sequence encoding, and self-attention mechanism for capturing the most informative subsequence. 
Sawhney et al.~\cite{sawhney2018exploring} used LSTM, CNN, and RNN for suicidal ideation detection. Similarly, Tadesse et al.~\cite{tadesse2020detection} employed LSTM-CNN model. Ji et al.~\cite{ji2020suicidal} proposed an attentive relation network with LSTM and topic modeling for encoding text and risk indicators. 

In the 2019 CLPsych Shared Task~\cite{zirikly2019clpsych}, many popular DNN architectures were applied. Hevia et al.~\cite{hevia2019analyzing} evaluated the effect of pretraining using different models, including GRU-based RNN. Morales et al.~\cite{morales2019investigation} studied several popular deep learning models such as CNN, LSTM, and Neural Network Synthesis (NeuNetS). Matero et al.~\cite{matero2019suicide} proposed dual-context model using hierarchically attentive RNN, and BERT. 

Another sub-direction is the so-called hybrid method, which cooperates minor feature engineering with representation learning techniques.
Chen et al.~\cite{chen2019similar} proposed a hybrid classification model of the behavioral model and the suicide language model. Zhao et al.~\cite{zhao2018text} proposed the D-CNN model taking word embedding and external tabular features as inputs for classifying suicide attempters with depression.

\begin{table*}[htp]
\scriptsize
\caption{Categorization of methods for suicidal ideation detection}
\begin{center}
\begin{tabular}{|c|c|c|c|}
\toprule
Category & Publications & Methods & Inputs \\
\midrule
\multirow{9}{1.5cm}{Feature Engineering} & Ji et al.~\cite{ji2018supervised} & Word counts, POS, LIWC, TF-IDF + classifiers & Text \\
				& Masuda et al.~\cite{masuda2013suicide} & Multivariate/univariate logistic regression & Characteristics variables\\
				& Delgado-Gomez et al.~\cite{delgado2012suicide} & International personal disorder examination screening questionnaire & Questionnaire responses\\
				& Mulholland et al.~\cite{mulholland2013suicidal} & Vocabulary features, syntactic features, semantic class features, N-gram & lyrics \\
				& Okhapkina et al.~\cite{okhapkina2017adaptation} & Dictionary, TF-IDF + SVD & Text\\
				& Huang et al.~\cite{huang2014detecting} & Lexicon, syntactic features, POS, tense & Text\\
				& Pestian et al.~\cite{pestian2010suicide} & Word counts, POS, concepts and readability score & Text \\
				& Tai et al.~\cite{tai2007artificial} & self-measurement scale + ANN & Self-measurement forms\\
				& Shing et al.~\cite{shing2018expert} & BoWs, empath, readability, syntactic, topic, LIWC, emotion, lexicon & Text \\
\midrule
\multirow{9}{1.6cm}{Deep Learning} & Zhao et al.~\cite{zhao2018text} & Word embedding, tabular features, D-CNN & Text+external information \\
				& Shing et al.~\cite{shing2018expert} & Word embedding, CNN, max pooling & Text \\
				& Ji et al.~\cite{ji2018supervised} & Word embedding, LSTM, max pooling& Text\\
				& Bento et al.~\cite{benton2017multi} & Multi-task learning, neural networks & Text \\
				& Nobles et al.~\cite{nobles2018identification} & MLP, psycholinguistic features, word occurrence & Text \\
				& Hevia et al.~\cite{hevia2019analyzing} & Pretrained GRU, word embedding, document embedding & Text \\
				& Morales et al.~\cite{morales2019investigation} & CNN, LSTM, NeuNetS, word embedding & Text \\
				& Matero et al.~\cite{matero2019suicide} & Dual-context, BERT, GRU, attention, user-factor adaptation & Text \\
				& Gaur et al.~\cite{gaur2019knowledge} & CNN, knowledge base, ConceptNet embedding & Text \\
				& Coppersmith et al.~\cite{coppersmith2018natural} & GloVe, BiLSTM, self attention & Text \\
				& Ji et al.~\cite{ji2020suicidal} & Relation network, LSTM, attention, lexicon & Text \\
				& Tadesse et al.~\cite{tadesse2020detection} & LSTM, CNN, word embedding & Text \\
\bottomrule
\end{tabular}
\end{center}
\label{tab:categorization}
\end{table*}%

\subsection{Summary}
The popularization of machine learning has facilitated research on suicidal ideation detection from multi-modal data and provided a promising way for effective early warning. Current research focuses on text-based methods by extracting features and deep learning for automatic feature learning. Researchers widely use many canonical NLP features such as TF-IDF, topics, syntactic, affective characteristics, and readability, and deep learning models like CNN and LSTM. Those methods, especially deep neural networks with automatic feature learning, boosted predictive performance and preliminary success on suicidal intention understanding. However, some methods may only learn statistical cues and lack of commonsense. The recent work~\cite{gaur2019knowledge} incorporated external knowledge using knowledge bases and suicide ontology for knowledge-aware suicide risk assessment. It took a remarkable step towards knowledge-aware detection. 

\section{Applications~on~Domains}
\label{sec:application}
\begin{figure*}[htbp]
\begin{center}
\begin{subfigure}[]{0.4\textwidth}
	\centering
	\includegraphics[height=5cm]{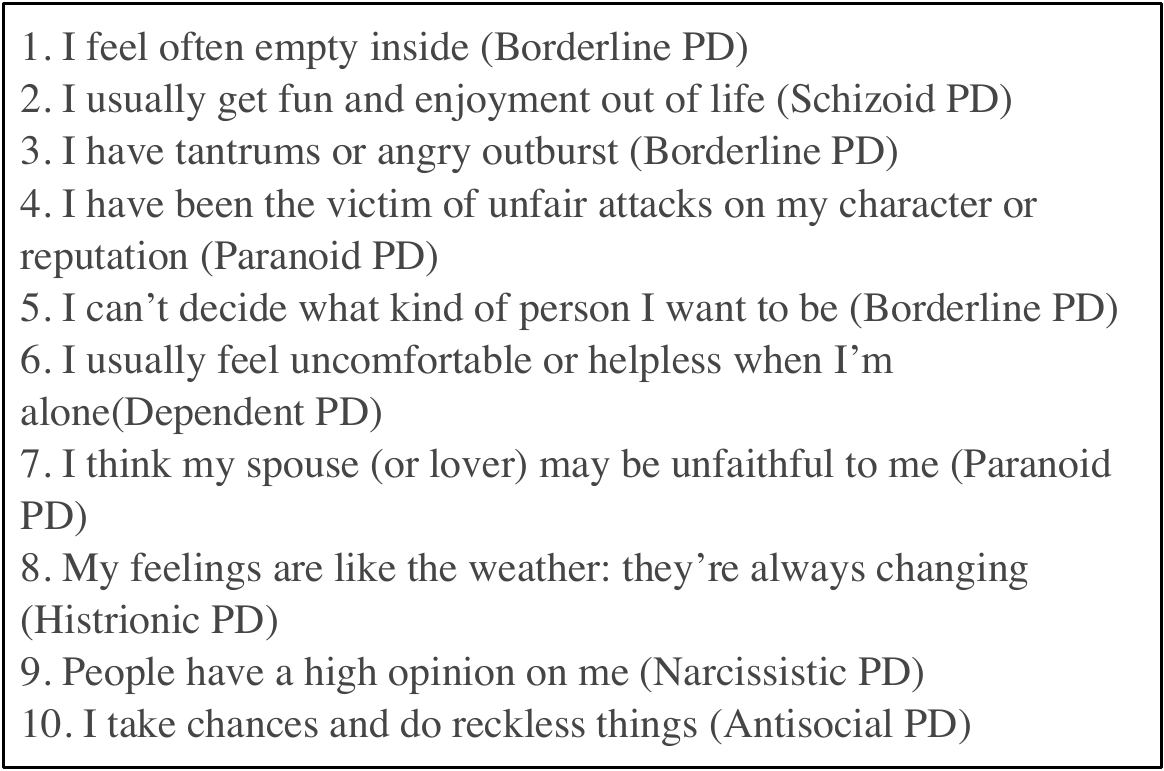}
	\subcaption{Questionnaire}
	\label{fig:questionnaire}
\end{subfigure}
\qquad
\begin{subfigure}[]{0.4\textwidth}
	\centering
	\includegraphics[height=5cm]{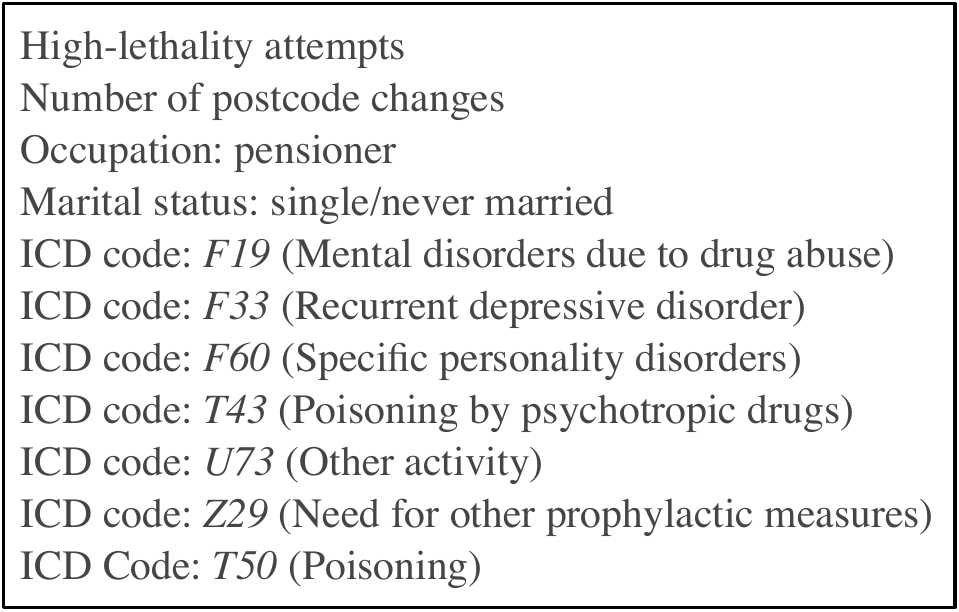}
	\subcaption{EHR}
	\label{fig:EHR}
\end{subfigure}
\qquad
\begin{subfigure}[]{0.4\textwidth}
	\centering
	\includegraphics[height=5cm]{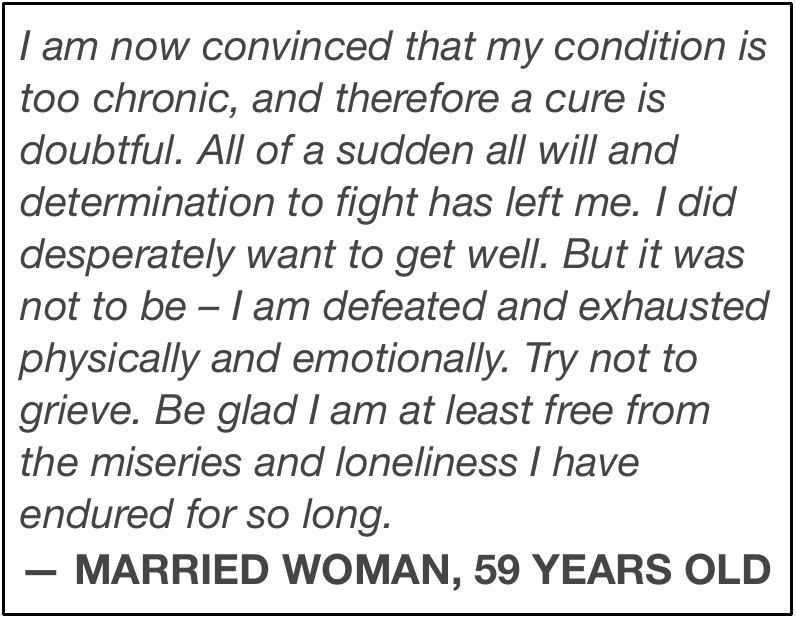}
	\subcaption{Suicide notes}
	\label{fig:notes}
\end{subfigure}
\qquad
\begin{subfigure}[]{0.4\textwidth}
	\centering
	\includegraphics[height=5cm]{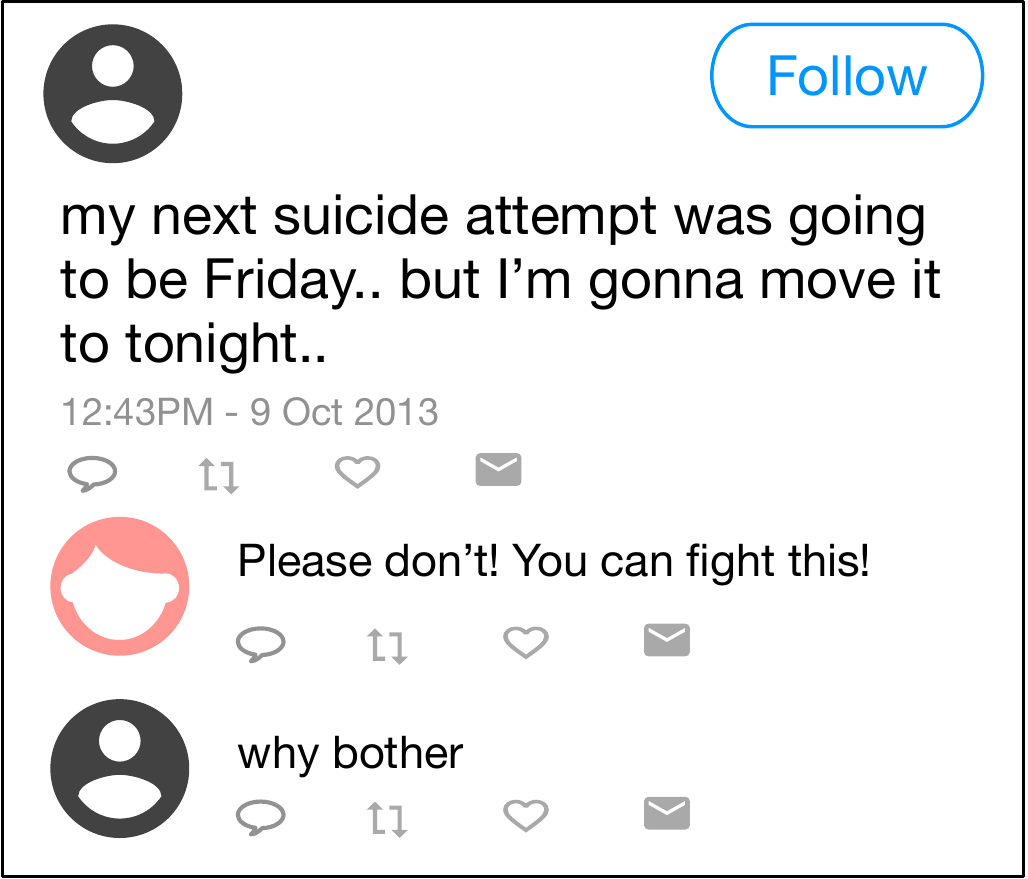}
	\subcaption{Tweets}
	\label{fig:tweets}
\end{subfigure}
\caption{Examples of content for suicidal ideation detection}
\label{fig:examples}
\end{center}
\end{figure*}

Many machine learning techniques have been introduced for suicidal ideation detection. The relevant extant research can also be viewed according to the data source. Specific applications cover a wide range of domains, including questionnaires, electronic health records (EHRs), suicide notes, and online user content. 
Fig.~\ref{fig:examples} shows some examples of data source for suicidal ideation detection, where Fig.~\ref{fig:questionnaire} lists selected questions of the ``International Personal Disorder Examination Screening Questionnaire'' (IPDE-SQ) adapted from~\cite{delgado2012suicide}, Fig.~\ref{fig:EHR} are selected patient's records from~\cite{tran2013integrated}, Fig.~\ref{fig:notes} is a suicide note from a website\furl{paranorms.com/suicide-notes}, and Fig.~\ref{fig:tweets} is a tweet and its corresponding comments from \url{Twitter.com}. Nobles et al.~\cite{nobles2018identification} identified suicide risk using text messages. Some researchers also developed softwares for suicide prevention. Berrouiguet et al.~\cite{berrouiguet2016toward} developed a mobile application for health status self report. Meyer et al.~\cite{meyer2017development} developed an e-PASS Suicidal Ideation Detector (eSID) tool for medical practitioners. Shah et al. \cite{shah2019multimodal} utilized social media videos and studied multimodal behavioral markers.

\subsection{Questionnaires}
\label{sec:questionnaires}
Mental disorder scale criteria such as DSM-IV\furl{psychiatry.org/psychiatrists/practice/dsm} and ICD-10\furl{apps.who.int/classifications/icd10/browse/2016/en}, and the IPDE-SQ provides good tool for evaluating an individual's mental status and their potential for suicide. Those criteria and examination metrics can be used to design questionnaires for self-measurement or face-to-face clinician-patient interview.

To study the assessment of suicidal behavior, Delgado-Gomez et al.~\cite{delgado2011improving} applied and compared the IPDE-SQ and the ``Barrat's Impulsiveness Scale''(version 11, BIS-11) to identify people likely to attempt suicide. The authors also conducted a study on individual items from those two scales. The BIS-11 scale has 30 items with 4-point ratings, while the IPDE-SQ in DSM-IV has 77 true-false screening questions. 
Further, Delgado-Gomez et al.~\cite{delgado2012suicide} introduced the ``Holmes-Rahe Social Readjustment Rating Scale'' (SRRS) and the IPDE-SQ as well to two comparison groups of suicide attempters and non-suicide attempters. The SRRS consists of 43 ranked life events of different levels of severity. Harris et al.~\cite{harris2014suicidal} surveyed understanding suicidal individuals' online behaviors to assist suicide prevention. Sueki~\cite{sueki2015association} conducted an online panel survey among Internet users to study the association between suicide-related Twitter use and suicidal behavior. Based on the questionnaire results, they applied several supervised learning methods, including linear regression, stepwise linear regression, decision trees, Lars-en, and SVMs, to classify suicidal behaviors. 

\subsection{Electronic~Health~Records}
\label{sec:EHR}
The increasing volume of electronic health records (EHRs) has paved the way for machine learning techniques for suicide attempter prediction. Patient records include demographical information and diagnosis-related history like admissions and emergency visits. However, due to the data characteristics such as sparsity, variable length of clinical series, and heterogeneity of patient records, many challenges remain in modeling medical data for suicide attempt prediction. Besides, the recording procedures may change because of the change of healthcare policies and the update of diagnosis codes.

There are several works of predicting suicide risk based on EHRs~\cite{hammond2013use, walsh2017predicting}. 
Tran et al.~\cite{tran2013integrated} proposed an integrated suicide risk prediction framework with a feature extraction scheme, risk classifiers, and risk calibration procedure. Explicitly, each patient's clinical history is represented as a temporal image. 
Iliou et al.~\cite{iliou2016machine} proposed a data preprocessing method to boost machine learning techniques for suicide tendency prediction of patients suffering from mental disorders. 
Nguyen et al.~\cite{nguyen2016evaluation} explored real-world administrative data of mental health patients from the hospital for short and medium-term suicide risk assessments. By introducing random forests, gradient boosting machines, and DNNs, the authors managed to deal with high dimensionality and redundancy issues of data. Although the previous method gained preliminary success, Iliou et al.~\cite{iliou2016machine} and Nguyen et al.~\cite{nguyen2016evaluation} have a limitation on the source of data which focuses on patients with mental disorders in their historical records. 
Bhat and Goldman-Mellor~\cite{bhat2017predicting} used an anonymized general EHR dataset to relax the restriction on patient's diagnosis-related history, and applied neural networks as a classification model to predict suicide attempters.

\subsection{Suicide Notes}
\label{sec:suicide-notes}
Suicide notes are the written notes left by people before committing suicide. They are usually written on letters and online blogs, and recorded in audio or video. 
Suicide notes provide material for NLP research. Previous approaches have examined suicide notes using content analysis~\cite{pestian2010suicide}, sentiment analysis~\cite{pestian2012sentiment, wang2012discovering}, and emotion detection~\cite{liakata2012three}. 
Pestian et al.~\cite{pestian2010suicide} used transcribed suicide notes with two groups of completers and elicitors from people who have a personality disorder or potential morbid thoughts. White and Mazlack~\cite{white2011discerning} analyzed word frequencies in suicide notes using a fuzzy cognitive map to discern causality. Liakata et al.~\cite{liakata2012three} employed machine learning classifiers to 600 suicide messages with varied length, different readability quality, and multi-class annotations. 

Emotion in text provides sentimental cues of suicidal ideation understanding. Desmet et al.~\cite{desmet2013emotion} conducted a fine-grained emotion detection on suicide notes of 2011 i2b2 task. Wicentowski and Sydes~\cite{wicentowski2012emotion} used an ensemble of maximum entropy classification. Wang et al.~\cite{wang2012discovering} and Kova{\v{c}}evi{\'c} et al.~\cite{kovavcevic2012topic} proposed hybrid machine learning and rule-based method for the i2b2 sentiment classification task in suicide notes. 

In the age of cyberspace, more suicide notes are now written in the form of web blogs and can be identified as carrying the potential risk of suicide. Huang et al.~\cite{huang2007hunting} monitored online blogs from \url{MySpace.com} to identify at-risk bloggers. Schoene and Dethlefs~\cite{schoene2016automatic} extracted linguistic and sentiment features to identity genuine suicide notes and comparison corpus. 

\subsection{Online User Content}
\label{sec:online}
The widespread use of mobile Internet and social networking services facilitates people's expressing their life events and feelings freely. As social websites provide an anonymous space for online discussion, an increasing number of people suffering from mental disorders turn to seek for help. There is a concerning tendency that potential suicide victims post their suicidal thoughts on social websites like Facebook, Twitter, Reddit, and MySpace. Social media platforms are becoming a promising tunnel for monitoring suicidal thoughts and preventing suicide attempts~\cite{robinson2016social}. Massive user-generated data provide a good source to study online users' language patterns. Using data mining techniques on social networks and applying machine learning techniques provide an avenue to understand the intent within online posts, provide early warnings, and even relieve a person's suicidal intentions.

Twitter provides a good source for research on suicidality. 
O'Dea et al.~\cite{odea2015detecting} collected tweets using the public API and developed automatic suicide detection by applying logistic regression and SVM on TF-IDF features. Wang et al.~\cite{wang2016role} further improved the performance with effective feature engineering. 
Shepherd et al.~\cite{shepherd2015using} conducted psychology-based data analysis for contents that suggests suicidal tendencies in Twitter social networks. The authors used the data from an online conversation called \#dearmentalhealthprofessionals. 

Another famous platform Reddit is an online forum with topic-specific discussions has also attracted much research interest for studying mental health issues~\cite{de2014mental} and suicidal ideation~\cite{huang2016online}. A community on Reddit called SuicideWatch is intensively used for studying suicidal intention~\cite{de2016discovering, ji2018supervised}. De Choudhury et al.~\cite{de2016discovering} applied a statistical methodology to discover the transition from mental health issues to suicidality. Kumar et al.~\cite{kumar2015detecting} examined the posting activity following the celebrity suicides, studied the effect of celebrity suicides on suicide-related contents, and proposed a method to prevent the high-profile suicides.

Many pieces of research~\cite{huang2014detecting, huang2015topic} work on detecting suicidal ideation in Chinese microblogs. Guan et al.~\cite{guan2015identifying} studied user profile and linguistic features for estimating suicide probability in Chinese microblogs. There also remains some work using other platforms for suicidal ideation detection. For example,
Cash et al.~\cite{cash2013adolescent} conducted a study on adolescents' comments and content analysis on MySpace.
Steaming data provides a good source for user pattern analysis. Vioul\`{e}s et al.~\cite{vioules2018detection} conducted user-centric and post-centric behavior analysis and applied a martingale framework to detect sudden emotional changes in the Twitter data stream for monitoring suicide warning signs. Ren et al.~\cite{ren2016examining} use the blog stream collected from public blog articles written by suicide victims to study the accumulated emotional information.

\subsection{Summary}
\label{sub:summary}
Applications of suicidal ideation detection mainly consist of four domains, i.e., questionnaires, electronic health records, suicide notes, and online user content. Table~\ref{tab:summary} gives a summary of categories, data sources, and methods. Among these four main domains, questionnaires and EHRs require self-report measurement or patient-clinician interactions and rely highly on social workers or mental health professions. Suicide notes have a limitation on immediate prevention, as many suicide attempters commit suicide in a short time after they write suicide notes. However, they provide a good source for content analysis and the study of suicide factors. The last online user content domain is one of the most promising ways of early warning and suicide prevention when empowered with machine learning techniques. With the rapid development of digital technology, user-generated content will play a more important role in suicidal ideation detection. Other forms of data, such as health data generated by wearable devices, can be very likely to help with suicide risk monitoring in the future.

 \begin{table}[!htbp]
 \small
 \caption{Summary of studies on suicidal ideation detection from the views of intervention categories, data and methods}
 \label{tab:summary}
 \centering
 \begin{tabular}{| c || l |}
 \toprule
 \multirow{3}{*}{Categories}& self-report examination~\cite{venek2017adolescent} \\
					& face-to-face suicide prevention~\cite{scherer2013investigating} \\
 					& automatic SID~\ref{sec:feature-engineering}~\ref{sec:deep-learning}~\ref{sec:suicide-notes}~\ref{sec:EHR}~\ref{sec:online}\\
 \midrule
 \multirow{5}{*}{Data} & questionnaires~\ref{sec:questionnaires} \\
 	& suicide notes~\ref{sec:suicide-notes} \\ 
 	& suicide blogs~\ref{sec:suicide-notes} \\
	& electronic health records~\ref{sec:EHR} \\
 	& online social texts~\ref{sec:online}\\
 \hline
 \multirow{5}{*}{Methods} & clinical methods~\cite{sikander2016predicting, just2017machine, jiang2015erp} \\
 					& mobile applications~\cite{tighe2017ibobbly} \\
 	 				& content analysis~\ref{sec:content-analysis} \\
 	 				& feature engineering~\ref{sec:feature-engineering}\\
					& deep learning~\ref{sec:deep-learning} \\
\hline
 \multirow{3}{*}{Critical issues} & suicide factors~\cite{hinduja2010bullying, joo2016death, vioules2018detection, ferrari2014burden, o2014psychology} \\
 					& ethics~\cite{de2018ethics, mckernan2018protecting, linthicum2019machine} \\
 					& privacy~\cite{de2018ethics} \\
 \bottomrule
 \end{tabular}
 \end{table}

\section{Tasks~and~Datasets}
\label{sec:tasks}
In this section, we summarize specific tasks in suicidal ideation detection and other suicide-related tasks about mental disorders. Some tasks, such as reasoning suicidal messages, generating a response, and suicide attempters detection on a social graph, may lack benchmarks for evaluation. However, they are critical for effective detection. We propose these tasks together with the current research direction and call for contribution to these tasks from the research community. Meanwhile, an elaborate list of datasets for currently available tasks is provided, and some potential data sources are also described to promote the research efforts. 

\subsection{Tasks}
\subsubsection{Suicide~Text~Classification}\label{sec:text-clf}
The first task - suicide text classification can be viewed as a domain-specific application of general text classification, which includes binary and multi-class classification. 
Binary suicidality classification simply determines text with suicidal ideation or not, while multi-class suicidality classification conducts fine-grained suicide risk assessment. For example, some research divides suicide risk into five levels: no, low, moderate, and severe. Alternatively, it can also consider four types of class labels according to mental and behavioral procedures, i.e., non-suicidal, suicidal thoughts/wishes, suicidal intentions, and suicidal act/plan.

Another subtask is risk assessment by learning from multi-aspect suicidal posts. Adopting the definition of characteristics of suicidal messages, Gilat et al.~\cite{gilat2011offering} manually tagged suicidal posts with multi-aspect labels, including mental pain, cognitive attribution, and level of suicidal risk.
Mental pain includes loss of control, acute loneliness, emptiness, narcissistic wounds, irreversibility loss of energy, and emotional flooding, scaled into $[0, 7]$. 
Cognitive attribution is the frustration of needs associated with interpersonal relationships, or there is no indication of attribution. 

\subsubsection{Reasoning~Suicidal~Messages}
Massive data mining and machine learning algorithms have achieved remarkable outcomes by using DNNs. However, simple feature sets and classification models are not predictive enough to detect complicated suicidal intentions. Machine learning techniques require reasoning suicidal messages to have a more in-depth insight into suicidal factors and the innermost being from textual posts. 
This task aims to employ interpretable methods to investigate suicidal factors and incorporate them with commonsense reasoning, which may improve the prediction of suicidal factors.
Specific tasks include automatic summarization of suicide factor, find an explanation of suicidal risk in mental pain and cognitive attribution aspects associated with suicide. 

\subsubsection{Suicide~Attempter~Detection}
The two tasks mentioned above focus on a single text itself. However, the primary purpose of suicidal ideation detection is the identify suicide attempters. Thus, it is vital to achieving user-level detection, which consists of two folds, i.e., user-level multi-instance suicidality detection and suicide attempt detection on a graph. The former takes a bag of posts from individuals as input and conducts multi-instance learning over a bag of messages. The later identifies suicide attempters in a specific social graph built by the interaction between users in social networks. It considers the relationship between social users and can be regarded as a node classification problem in a graph.

\subsubsection{Generating~Response}
The ultimate goal of suicidal ideation detection is intervention and suicide prevention. Many people with suicidal intentions tend to post their suffering at midnight. Another task is generating a thoughtful response for counseling potential suicidal victims to enable immediate social care and relieve their suicidal intention. 
Gilat et al.,~\cite{gilat2011offering} introduced eight types of response strategies; they are emotional support, offering group support, empowerment, interpretation, cognitive change inducement, persuasion, advising, and referring. This task requires machine learning techniques, especially sequence-to-sequence learning, to have the ability to adopt effective response strategies to generate better response and eliminate people's suicidality. 
When social workers or volunteers go back online, this response generation technique can also generate hints for them to compose a thoughtful response.

\subsubsection{Mental~disorders~and~Self-harm~risk}
Suicidal ideation has a strong relationship with a mental health issue and self-harm risks. Thus, detecting severe mental disorders or self-harm risks is also an important task.
Such works include depression detection~\cite{yates2017depression}, self-harm detection~\cite{wang2017understanding}, stressful periods and stressor events detection~\cite{li2016analyzing}, building knowledge graph for depression~\cite{huang2017constructing}, and correlation analysis on depression and anxiety~\cite{hao2019providing}. Corresponding subtasks in this field are similar to suicide text classification in Section~\ref{sec:text-clf}. 

\subsection{Datasets}
\subsubsection{Text~Data}
\paragraph{Reddit} 
Reddit is a registered online community that aggregates social news and online discussions. It consists of many topic categories, and each area of interest within a topic is called a subreddit. A subreddit called ``Suicide Watch''(SW)\furl{reddit.com/r/SuicideWatch} is intensively used for further annotation as positive samples. Posts without suicidal content are sourced from other popular subreddits. Ji et al.~\cite{ji2018supervised} released a dataset with 3,549 posts with suicidal ideation. Shing et al.~\cite{shing2018expert} published their UMD Reddit Suicidality Dataset with 11,129 users and total 1,556,194 posts and sampled 934 users for further annotation. Alada{\u{g}} et al.~\cite{aladaug2018detecting} collected 508,398 posts using Google Cloud BigQuery, and manually annotated 785 posts. 

\paragraph{Twitter}
Twitter is a popular social networking service, where many users also talk about their suicidal ideation. Twitter is quite different from Reddit in post length, anonymity, and the way communication and interaction.
Twitter user data with suicidal ideation and depression are collected by Coppersmith et al.~\cite{coppersmith2015quantifying}. Ji et al.~\cite{ji2018supervised} collected an imbalanced dataset of 594 tweets with suicidal ideation out of a total of 10,288 tweets. 
Vioul{\`e}s et al. collected 5,446 tweets using Twitter streaming API~\cite{vioules2018detection}, of which 2,381 and 3,065 tweets from the distressed users and normal users, respectively.
However, most Twitter-based datasets are no longer available as per the policy of Twitter. 

\paragraph{ReachOut}
ReachOut Forum\furl{au.reachout.com/forums} is a peer support platform provided by an Australian mental health care organization.
The ReachOut dataset~\cite{milne2016clpsych} was firstly released in the CLPsych17 shared task.
Participants were initially given a training dataset of 65,756 forum posts, of which 1188 were annotated manually with the expected category, and a test set of 92,207 forum posts, of which 400 were identified as requiring annotation. The specific four categories are described as follows. 
1) crisis: the author or someone else is at risk of harm; 2) red: the post should be responded to as soon as possible; 3) amber: the post should be responded to at some point if the community does not rally strongly around it; 4) green: the post can be safely ignored or left for the community to address.

\subsubsection{EHR}
EHR data contain demographical information, admissions, diagnostic reports, and physician notes. 
A collection of electronic health records is from the California emergency department encounter and hospital admission. It contains 522,056 anonymous EHR records from California-resident adolescents. However, it is not public for access. 
Bhat and Goldman-Mellor~\cite{bhat2017predicting} firstly used these records from 2006-2009 to predict the suicide attempt in 2010.
Haerian et al.~\cite{haerian2012methods} selected 280 cases for evaluation from the Clinical Data Warehouse (CDW) and WebCIS database at NewYork Presbyterian Hospital/ Columbia University Medical Center. 
Tran et al.~\cite{tran2013integrated} studied emergency attendances with a least one risk assessment from the Barwon Health data warehouse. The selected dataset contains 7,746 patients and 17,771 assessments. 

\subsubsection{Mental~Disorders}
Mental health issues such as depression without effective treatment can turn into suicidal ideation. For the convenience of research on mental disorders, we also list several resources for monitoring mental disorders.
The eRisk dataset of Early Detection of Signs of Depression~\cite{losada2016test} is released by the first task of the 2018 workshop at the Conference and Labs of the Evaluation Forum (CLEF), which focuses on early risk prediction on the Internet\furl{early.irlab.org}. This dataset contains sequential text from social media.
Another dataset is the Reddit Self-reported Depression Diagnosis (RSDD) dataset~\cite{yates2017depression}, which contains 9,000 diagnosed users with depression and approximately 107,000 matched control users.

\begin{table*}[htp]
\small
\caption{A summary of public datasets}
\begin{center}
\begin{tabular}{|c|c|c|c|c|}
\toprule
Type & Publication & Source & Instances & Public Access \\
\midrule
Text & Shing et al.~\cite{shing2018expert} & Reddit & 866/11,129 & \url{https://tinyurl.com/umd-suicidality}\\
Text & Alada{\u{g}} et al.~\cite{aladaug2018detecting} & Reddit & 508,398 & Request to the authors \\
Text & Coppersmith et al.~\cite{coppersmith2015quantifying} & Twitter & $>$ 3,200& N.A \\
Text & Coppersmith et al.~\cite{coppersmith2015clpsych} & Twitter & 1,711 & \url{https://tinyurl.com/clpsych-2015} \\
Text & Vioul{\`e}s et al.~\cite{vioules2018detection} & Twitter & 5,446 & N.A \\
Text & Milne et al.~\cite{milne2016clpsych} & ReachOut & 65,756 & N.A \\
EHR & Bhat and Goldman-Mellor~\cite{bhat2017predicting} & Hospital & 522,056 & N.A\\
EHR & Tran et al.~\cite{tran2013integrated} & Barwon Health & 7,746 & N.A \\
EHR & Haerian et al.~\cite{haerian2012methods} & CDW \& WebCIS & 280 & N.A \\
Text & Pestian et al.~\cite{pestian2012sentiment} & Notes & 1,319 & 2011 i2b2 NLP challenge \\
Text & Gaur et al.~\cite{gaur2019knowledge} & Reddit & 500/2,181 & Request to the authors \\
Text & eRisk 2018~\cite{losada2016test} & Social networks & 892 & {\url{https://tec.citius.usc.es/ir/code/eRisk.html}} \\
Text & RSDD~\cite{yates2017depression} & Reddit & 9,000 & {\url{http://ir.cs.georgetown.edu/resources/rsdd.html}} \\
\bottomrule
\end{tabular}
\end{center}
\label{tab:datasets}
\end{table*}%

\section{Discussion~and~Future~Work}
\label{sec:discussion}
Many preliminary works have been conducted for suicidal ideation detection, especially boosted by manual feature engineering and DNN-based representation learning techniques. However, current research has several limitations, and there are still great challenges for future work. 

\subsection{Limitations}
\paragraph{Data~Deficiency}
The most critical issue of current research is data deficiency. Current methods mainly apply supervised learning techniques, which require manual annotation. However, there are not enough annotated data to support further research. For example, labeled data with fine-grained suicide risk only have limited instances, and there are no multi-aspect data and data with social relationships. 

\paragraph{Annotation~Bias}
There is little evidence to confirm the suicide action to obtain ground truth. Thus, current data are obtained by manual labeling with some predefined annotation rules. The crowdsourcing-based annotation may lead to bias of labels. Shing et al.~\cite{shing2018expert} asked experts for labeling but only obtained a limited number of labeled instances. 
As for the demographical data, the quality of suicide data is concerning, and mortality estimation is general death but not suicide\footnote{World Health Organization, Preventing suicide: a global imperative, 2014. \url{https://apps.who.int/iris/bitstream/handle/10665/131056/9789241564779_eng.pdf}}. Some cases are misclassified as accidents or death of undetermined intent. 

\paragraph{Data~Imbalance}
Posts with suicidal intention account for a tiny proportion of massive social posts. However, most works built datasets in an approximately even manner to collect relatively balanced positive and negative samples rather than treating it as an ill-balanced data distributed. 

\paragraph{Lack of Intention Understanding}
The current statistical learning method failed to have a good understanding of suicidal intention. 
The psychology behind suicidal attempts is complex. However, mainstream methods focus on selecting features or using complex neural architectures to boost the predictive performance. 
From the phenomenology of suicidal posts in social content, machine learning methods learned statistical clues. However, they failed to reason over the risk factors by incorporating the psychology of suicide. 

\subsection{Future~Work}
\subsubsection{Emerging~Learning~Techniques}
The advances of deep learning techniques have boosted research on suicidal ideation detection. More emerging learning techniques, such as attention mechanism and graph neural networks, can be introduced for suicide text representation learning. Other learning paradigms, such as transfer learning, adversarial training, and reinforcement learning, can also be utilized. For example, knowledge of the mental health detection domain can be transferred for suicidal ideation detection, and generative adversarial networks can be used to generated adversarial samples for data augmentation. 

In social networking services, posts with suicidal ideation are in the long tail of the distribution of different post categories. To achieve effective detection in the ill-balanced distribution of real-world scenarios, few-shot learning can be utilized to train on a few labeled posts with suicidal ideation among the large social corpus. 
						
\subsubsection{Suicidal~Intention~Understanding~and~Interpretability}
Many factors are correlated to suicide, such as mental health, economic recessions, gun prevalence, daylight patterns, divorce laws, media coverage of suicide, and alcohol use\footnote{Report by Lindsay Lee, Max Roser, and Esteban Ortiz-Ospina in OurWorldInData.org, retrieved from \url{https://ourworldindata.org/suicide}}. 
A better understanding of suicidal intention can provide a guideline for effective detection and intervention. A new research direction is to equip deep learning models with commonsense reasoning, for example, by incorporating external suicide-related knowledge bases. 

Deep learning techniques can learn an accurate prediction model. However, this would be a black-box model. In order to better understand people's suicidal intentions and have a reliable prediction, new interpretable models should be developed.

\subsubsection{Temporal~Suicidal~Ideation~Detection}
Another direction is to detect suicidal ideation over the data stream and consider the temporal information. There exist several stages of suicide attempts, including stress, depression, suicidal thoughts, and suicidal plan. Modeling people's posts' temporal trajectory can effectively monitor the change of mental status and is essential for detecting early signs of suicidal ideation.

\subsubsection{Proactive~Conversational~Intervention}
The ultimate aim of suicidal ideation detection is intervention and prevention. Very little work is undertaken to enable proactive intervention. Proactive Suicide Prevention Online (PSPO)~\cite{liu2019proactive} provides a new perspective with the combination of suicidal identification and crisis management. An effective way is through conversations. Automatic response generation becomes a promising technical solution to enable timely intervention for suicidal thoughts. Natural language generation techniques can be utilized to generate counseling responses to comfort people's depression or suicidal ideation. 
Reinforcement learning can also be applied for conversational suicide intervention. 
After suicide attempters post suicide messages (as the initial state), online volunteers and lay individuals will take action to comment on the original posts and persuade attempters to give up their suicidality. The attempter may do nothing, reply to the comments, or get their suicidality relieved. A score will be defined by observing the reaction from a suicide attempter as a reward. The conversational suicide intervention uses a policy gradient for agents to generated responses with maximum rewards to best relieve people's suicidal thoughts. 

\section{Conclusion}
Suicide prevention remains an essential task in our modern society. Early detection of suicidal ideation is an important and effective way to prevent suicide. This survey investigates existing methods for suicidal ideation detection from a broad perspective which covers clinical methods like patient-clinician interaction and medical signal sensing; textual content analysis such as lexicon-based filtering and word cloud visualization; feature engineering including tabular, textual, and affective features; and deep learning-based representation learning like CNN- and LSTM-based text encoders. Four main domain-specific applications on questionnaires, EHRs, suicide notes, and online user content are introduced. 

Psychological experts have conducted most work in this field with statistical analysis, and computer scientists with feature engineering based machine learning and deep learning-based representation learning. Based on current research, we summarized existing tasks and further proposed new possible tasks. Last but not least, we discuss some limitations of current research and propose a series of future directions, including utilizing emerging learning techniques, interpretable intention understanding, temporal detection, and proactive conversational intervention. 

Online social content is very likely to be the main channel for suicidal ideation detection in the future. Therefore, it is essential to develop new methods, which can heal the schism between clinical mental health detection and automatic machine detection, to detect online texts containing suicidal ideation in the hope that suicide can be prevented.

\bibliographystyle{IEEEtran}

\end{document}